\documentstyle[12pt,titlepage]{article}
\input epsfig.sty

\renewcommand{\theequation}{\thesection.\arabic{equation}}

\font\grande=cmr10 scaled \magstep4
\font\medio=cmr10 scaled \magstep2
\outer\def\beginsection#1\par{\medbreak\bigskip
      \message{#1}\leftline{\bf#1}\nobreak\medskip\vskip-\parskip
      \noindent}

\def\laq{\raise 0.4ex\hbox{$<$}\kern -0.8em\lower 0.62
ex\hbox{$\sim$}}
\def\gaq{\raise 0.4ex\hbox{$>$}\kern -0.7em\lower 0.62
ex\hbox{$\sim$}}

\begin{document}
\bibliographystyle{unsrt}
\titlepage
\begin{center}
{\grande Gravitational Waves Constraints}\\
\vspace{5mm}
{\grande on Post-Inflationary Phases Stiffer than Radiation}\\
\vspace{15mm}
\centerline{Massimo Giovannini \footnote{Electronic address:
m.giovannini@damtp.cam.ac.uk}}
\bigskip
\centerline{{\em DAMTP, 21 Silver Street, Cambridge, CB3 9EW, United
Kingdom}}
\smallskip
\end{center}
\vspace{10mm}
\centerline{\medio  Abstract}
\noindent
We point out that the existence of post-inflationary phases stiffer
than radiation leads to the production of stochastic gravitational
waves (GW) backgrounds whose logarithmic energy spectra (in critical
units) are typically ``blue" at high frequencies. The maximal
spectral slope (for present frequencies larger than $10^{-16}$ Hz) is
of order one and it is related to the maximal sound velocity of the
stiff plasma governing the evolution of the geometry.
The duration of the stiff phase is crucially determined by the
backreaction of the GW leaving the horizon during the de Sitter phase
and re-entering during the stiff phase. Therefore, the maximal
(inflationary) curvature scale has to be fine-tuned to a value
smaller than the  limits set by the large scale measurements
($H_{dS}\laq 10^{-6}~M_{P}$) in order to a have a sufficiently long
stiff phase reaching an energy scale of the order of the $1~{\rm
TeV}$ and even lower if we want the  stiff phase to touch the
hadronic era (corresponding to  $T_{had}\sim 140~{\rm MeV}$).
By looking more positively at our exercise we see that, if an
inflationary phase is followed by a stiff phase, there exist an
appealing possibility of ``graviton reheating" whose effective
temperature can be generally quite low. \newpage

\renewcommand{\theequation}{1.\arabic{equation}}
\setcounter{equation}{0}
\section{Introduction}

Strong causality arguments \cite{kolbandturner}
 forbid the existence of a never ending radiation
dominated epoch. If this would be the case
regions emitting a highly homogenous and isotropic Cosmic Microwave
Background Radiation (CMBR) at the decoupling epoch, would not have
been
in causal contact in the far past. This problem of horizons (together
with
other kinematical problems of the standard cosmological model) led to
assume an inflationary phase \cite{inflation}
 of accelerated expansion ($\dot{a} >0$, $\ddot{a}>0$) where the
(effective)
equation of state describing the  background sources during inflation
 was drastically deviating from the one of radiation.

In the context of ordinary inflationary models \cite{inflation} the
transition from the inflationary regime to the radiation era is
usually associated with a reheating phase where the energy density of
the inflaton field is released, ``producing" a radiation dominated
phase. The dynamics of the inflaton right after the inflationary
epoch has been recently discussed in detail \cite{preheating}.

If this if the dynamical picture of the evolution of our Universe in
its early stages, then, one of the most interesting (and difficult to
test) implications is the production of a stochastic background of
gravitational waves (GW). It is indeed known since many years that
the various transitions of the curvature scale lead, necessarily, to
the amplifications of the quantum mechanical (vacuum) fluctuations of
the tensor modes of the geometry \cite{grishchuk2} and to the
consequent production of highly coorelated graviton squeezed states
\cite{gris4,max2}.
According to this mechanism, the energy density of gravitational
origin can be estimated to be (today) of the order of $10^{-13}$
\cite{all, sahni, grishchuk4} (in critical units) for (present)
frequencies larger than $10^{-16}$ Hz. The (logarithmic) energy
spectrum turns out to be flat in the same interval of frequencies.
The reason of this quite minute amplitude comes essentially from
the measurement of the CMBR anisotropies.
In fact, the tensor contribution to the CMBR anisotropies imposes a
quite
important bound on the maximal curvature scale at which the
inflationary
expansion occurred. By assuming, for example, that the inflationary
phase was of de Sitter type, with typical curvature scale $H_{dS}$,
 we have to require \cite{rubakov,aniso},
\begin{equation}
\frac{H_{dS}}{M_{P}} \laq 10^{-6},
\label{bound}
\end{equation}
in order to be compatible with the detected level of anisotropies in
the
microwave sky. Constraints on the dynamical assumptions of various
inflationary models can be derived on the same basis
\cite{grishchuk1}.

In this paper we want to explore a slightly different picture of the
post-inflationary phase. Our suggestion is, in short, the following.
Suppose that the inflationary phase is not immediately followed by a
radiation dominated phase but by an intermediate phase whose equation
of state is stiffer than radiation (i.e. $p= \gamma \rho$, with
$\gamma>1$ and $\gamma= c^2_{s}$  [$c_{s}$ is the sound velocity of
the plasma]). Then, two interesting implications can arise. On one
hand the transition between the inflationary regime and the stiff
regime leads to graviton spectra which slightly increase with
frequency (with ``blue" slopes) in the ultra-violet branch of the
spectrum, on the other hand, the backreaction effects associated with
the GW leaving the horizon during the de Sitter phase and re-entering
during the stiff phase can heat up the Universe very efficiently,
leading to a qualitatively new kind of reheating which one can call
``graviton reheating".

Even if the implications of this suggestion can be appealing it is
certainly important to justify better some possible motivations of
such a weird exercise.

First of all we can say that such a suggestion is not forbidden by
any present data.
An indirect evidence of the fact that the Universe might have
been  dominated
by radiation around temperatures of the order of $0.1$
MeV comes from the success of the simplest
(homogeneous and isotropic) big-bang nucleosynthesis (BBN) scenario
\cite{ns}.
In the absence of any external magnetic field \cite{magnetic}
and of matter--antimatter domains \cite{misha1}, the light
elements ($^3$ He, $^4$ He, Li, D) abundances are reproduced by the
BBN model provided the ratio of the baryonic charge density over
 the photon density is fine tuned around $10^{-10}$.
However, prior to nucleosynthesis, there are no direct tests of the
thermodynamical state of the Universe and, therefore, the effective
equation of state of the perfect fluid sources driving the evolution
of the background geometry can be arbitrarily different from the one
of a radiation dominated plasma. Moreover, there are no compelling
reasons why long-range (Abelian) gauge fields should not have been
present in the early Universe \cite{misha1}. Owing to our ignorance
of the thermodynamical state of the Universe
prior to the nucleosynthesis epoch, it is  possible to
postulate, in the framework of a particular model, the existence of
post-inflationary (decelerated) phases different from radiation.

The possibility of having post-inflationary phases whose (effective)
equation of state was stiffer than radiation can be also motivated in
terms of different models. It was recently argued that  a stiff
phase could be originated by the relaxation of the moduli towards the
minimum of their non-perturbative potential \cite{banks}. A similar
idea was investigated in the framework of the dilaton relaxation
 \cite{max} where the resonant
amplification of gauge fields modes was also discussed.
In fact
one can argue that it does certainly exist a regime where the kinetic
energy of the single modulus is dominant against its potential
energy. Therefore, the geometry would evolve following an effective
equation of state whose sound velocity is approximately equal to the
speed of light. In this regime the energy density of the modulus
would
scale as $\rho_{mod}\sim a^{-6}$. Of course intermediate
situations can be also imagined so that we can, say, more generally
that some phases with $ 1/3 < \gamma \leq 1$ can occur as a result of
the moduli relaxation. For
 example, in the context of dilaton relaxation \cite{max},
 if the kinetic energy dominates we get exactly a stiff fluid
 model with $\gamma=1$.
  In fact, on theoretical ground the dilaton
 field ($\phi$) has a
 potential  going  to zero (in the supersymmetric limit) as a double
 exponential (i.e. $V \sim \exp{[ -c^2 \exp{( -\phi)}]}$, with $c^2$
 positive and model dependent constant). On more physical ground the
 $V(\phi)$ is
 believed to have one (or more) minima for some $ \phi\sim
 \phi_{\min}$. When $\phi$ reaches
 curvature scale $H \sim m$ (where $m$ is the
 dilaton mass) an oscillating phase begins. Depending upon the
initial
 conditions in $H_1$ (the maximal curvature)
 the oscillating phase can be preceded by a stiff  phase where the
 dilaton kinetic energy dominates ($\dot{\phi}^2 \gg V(\phi)$),
and, therefore the evolution of the
 dilaton, in curvature, will be $\phi = \phi_{0} + \phi_{1}
 \log{[H/H_1]}$ implying $\dot{\phi}^2 \sim a^{-6}$.

It was recently speculated \cite{ew} (without relation to moduli
relaxation) that, provided the stiff phase is long enough,
interesting effects can also be expected at energy scales of the
order of $100$ GeV (corresponding to a curvature scale of the order
$H_{ew}\sim 10^{-34} M_{P}$).
Historically, the first one to imagine the appealing possibility of
having long stiff epochs was
Zeldovich \cite{zeldovich}. At that time inflationary models were not
yet formulated and it seemed quite crucial to model correctly  the
quark-hadron phase transition occurring when the temperature of the
Universe was of the order of the pion mass
(i.e.  $T_{had} \sim 140~{\rm MeV}$). The idea
was that, at this stage, the equation of state of the perfect fluid
sources could  be stiffer than the one of radiation,
namely with $\gamma>1/3$. On
physical ground, one can easily understand that the sound velocity in
a
perfect fluid is likely to be smaller than the speed of light
($\gamma\leq 1$ in our units). In particular the stiff model of
Zeldovich assumes exactly that, prior to the hadronic stage of
evolution, $\gamma=1$.
The fact that the speed of light equals the speed of sound also
implies  that the energy density of the stiff sources decreases
faster
than radiation.

This last feature of the stiff picture of hadronic interactions
 is  indeed fatal for the logical consistency of the whole proposal.
Actually, the existence of a hadronic phase with effective equation
of state stiffer than radiation also implies the production of a
stochastic background of GW  sharply peaked towards
 the Planck frequency \cite{grishchuk3}. Thus, the produced
gravitational waves
 will backreact on the geometry (effectively driven by the stiff
 fluid). Now, since the energy density of the high frequency GW
re-entering the
 horizon during the stiff epoch scales like radiation
 \cite{grishchuk3, brandenberger} it was
 correctly concluded \cite{grishchuk3} that, if a stiff phase ever
 existed prior to the usual radiation dominated phase, the
 back reaction effects associated with the production of high
frequency
 gravitons turned very quickly the evolution of the Universe into a
radiation
 dominated phase. It was also  showed
\cite{grishchuk3,parker} that there are no chances of having a stiff
 phase from the Planck curvature scale ($H_{P} \sim M_{P}$) down to
the
 hadronic curvature scale ($H_{had} \sim 10^{-40} M_{P}$).

In our context there is a crucial difference with respect to
Zeldovich suggestion and it is essentially given by Eq.
(\ref{bound}). Since the curvature scale is quite minute in Planck
units GW backreaction is not  switched on immediately.
None the less, the aim of this exercise is to point out that the
occurrence
(and the duration) of a post-inflationary phase stiffer than
radiation is not a
free parameter which we can adjust in the framework of a particular
model to get the desired effects. On the contrary the duration of a
stiff phase is
significantly constrained by the back reaction of hard gravitons.
In this sense our considerations owe very much to the
pioneering works in the subject \cite{grishchuk3,parker} but are
applied to a different dynamical picture, where an inflationary phase
is followed by a stiff phase. Our logic, in short, is the
following. Let us assume, for example that in some specific model a
de
Sitter (inflationary) phase is followed by a stiff phase (for example
with $p = \rho$) at some cosmic time $t_{1}$.
Then, the hard gravitons excited during the de
Sitter phase will have typical (physical) momentum
$\hbar \omega_{dS}(t_{1}) \sim \hbar H_{dS}(t_{1})$. Since the the
Hubble distance $H^{-1}$ deviates (during the stiff phase) from the
value during the de Sitter epoch as a consequence of the change in
the
expansion rate the hard gravitons with $\omega \leq \omega_{dS}$ will
``re-enter'' at different times during the stiff phase. Since their
effective equation of state is the one of radiation they will modify
the dynamics of the stiff phase by ultimately destroying it and by
turning it into a radiation dominated phase. By looking positively at
this effect we could say that the graviton back reaction represents a
reasonable candidate in order to implement a reasonable reheating
mechanism in these classes of models. At the same time the
constraints
imposed by the (over)-production of gravitons might be viewed as a
weakness of the scenario. A related result  of our analysis
will be the calculation of the GW spectra produced
during these types of stiff post-inflationary phases.

The plan of our paper is then the following. In Sec. II we will
introduce the basic equations describing the background
evolution during a stiff, post-inflationary, phase.  Sec. III is
devoted to the calculation of the GW spectra in these
models. In Sec. IV we will discuss the back reaction effects
associated
with hard (non-thermal) gravitons and we will discuss the possible
modification of the background evolution. Sec. IV contains our
concluding remarks.

\renewcommand{\theequation}{2.\arabic{equation}}
\setcounter{equation}{0}
\section{Basic Equations}

In this Section we consider the simplest homogeneous and isotropic
models of FRW type with line element
\begin{equation}
ds^2 = g_{\mu\nu} dx^{\mu} dx^{\nu}= dt^2 - a^2(t) \biggl[
\frac{dr^2}{1 - {\kappa}r^2} + r^2 ( d\theta^2 + \sin^2{\theta}
d\phi^2)\biggr]
\label{line}
\end{equation}
(Greek indices run from $0$ to $3$, whereas Latin indices run from
$1$
to $3$). The
derivative with respect to the cosmic time $t$ will be denoted by an
over dot whereas the derivative with respect to the conformal time
$\eta$ will be denoted by a prime (as usual $a(\eta) d \eta =
dt$). The sign of the spatial curvature $\kappa = +1,0,-1$
corresponds
to a closed, flat or open space. We will assume that the evolution of
the geometry follows general relativity. Therefore the coupled
evolution of the (perfect) fluid sources and of the
geometry will be conveniently
described in terms of the well known FRW equations
\begin{eqnarray}
M^2_{P} \biggl[ H^2  - \frac{k}{a^2}\biggr] &=& \rho -
\frac{\kappa}{a^2},
\nonumber\\
M^2_{P}\biggl[ H^2 + \dot{H}\biggr] &=& - \frac{1}{2}(\rho + 3 p),
\nonumber\\
\dot{\rho} + 3 H( \rho + p) &=& 0,~~~H= \frac{ d \log{a}}{dt},
\label{FRW}
\end{eqnarray}
where $H$ is the Hubble parameter and we also write
$M_{P}= \biggl(8\pi G/3\biggr)^{-1/2}$. Focusing our attention
on the conformally flat case (${\kappa} =0$)
we will consider models of background evolution where an inflationary
phase ($\dot{a}>0$, $\ddot{a} >0$)
 is followed at some time $t_1$ by a decelerated phase ($\ddot{a}<0$,
 $\dot{a}>0$). In particular we will assume that the sources for
$t>t_1$
will be well approximated by a barotropic equation of state
\begin{equation}
p= \gamma \rho,
\label{state}
\end{equation}
with $ 1/3 <\gamma \leq 1$. We consider unrealistic the case where
$\gamma >1$ since this would mean that the sound velocity of the
fluid
 is greater than the speed of light.
By integrating the FRW equations (\ref{FRW}) after $t_1$ with the
closure given by Eq. (\ref{state})  we get that
\begin{equation}
a(t) = a_{1} \biggl(\frac{t}{\alpha t_1}\biggr)^{\alpha},~~~\rho(t) =
\rho_{1} \biggl(\frac{a_{1}}{a}\biggr)^{\frac{2}{\alpha}}
\label{stiffsol}
\end{equation}
with
\begin{equation}
\alpha= \frac{2}{3(\gamma+1)}, ~~~\rho_{1} = H^2_{1}M^2_{P},~~~ a_1=
a(t_{1}),
\end{equation}
where $H_{1}= H(t_1)$ is simply the value of the Hubble
parameter at the end of the inflationary phase.

We want to stress that our
 approach in the present section will be an effective one: since
the evolution equations of the tensor modes of the geometry is
essentially determined {\em only} by the behavior of the scalar
curvature we feel free of using the simplest fluid model for the
description of the background sources. This is of course not a
limitation. The effective fluid sources can be thought as modelled by
 the energy momentum tensor of one (or more) scalar fields.

Concerning the inflationary phase we will not make any type of weird
assumption about the background evolution. Indeed we will show
that our considerations will be only mildly sensitive to the specific
inflationary dynamics and will only (but crucially) depend upon the
maximal curvature scale reached during inflation.
{}From a purely kinematical point of
view we will explore expanding inflationary epochs (i.e.
 $\dot{a}>0$, $\ddot{a} >0$) with constant or decreasing curvature
(i.e. $\dot{H}\leq 0$). The de Sitter (or quasi de Sitter)
 case seem to emerge quite naturally in
the framework of the slow-rolling
approximation\cite{inflation}. It is important to point out that the
de Sitter
case and the power-law case will produce, respectively, either flat
or
decreasing energy spectra \cite{grishchuk1,all,sahni,grishchuk4}.

Having said this we will focus our attention to the case where a pure
de Sitter phase is followed by a stiff phase and we will comment,
where appropriate, on the other possible cases.
Therefore the model we want to investigate is given
(in conformal time)  essentially by
three phases:
\begin{eqnarray}
a_{i}(\eta) &=& -\biggl(\frac{\eta_{1}}{\eta}\biggr),~~~~\eta<
-\eta_{1}
\nonumber\\
a_{s}(\eta) &=& \biggl[\frac{ (1+ \beta) \eta_{1} +
\eta}{\beta\eta_1}\biggr]^{\beta},~~~~-\eta_{1}<\eta<\eta_{r}
\nonumber\\
a_{r}(\eta) &=& \frac{\beta \eta + (\beta + 1)\eta_1 - (\beta -1)
\eta_{r} }{[ \beta \eta_{1}]^{\beta} [
\eta_{r} + (\beta+1) \eta_{1}]^{1 - \beta}}, ~~~~\eta>\eta_{r}.
\label{model}
\end{eqnarray}
The  subscript  $i, s$ and $r$ simply stand for
inflationary, stiff and radiation dominated phases. An important
feature of Eq. (\ref{model}) is the continuity of the  scale factors
(and of their first conformal time derivatives) in the matching
points
$\eta_{1}$ and $\eta_{r}$. Notice also that the generic exponent
$\beta $ specifying the dynamics during the stiff phase is trivially
related to the, previously introduced $\alpha$ and $\gamma$
parameters, namely
\begin{equation}
\beta = \frac{\alpha}{1-\alpha} = \frac{2}{3 \gamma + 1}.
\end{equation}
Of course the three phases specified by Eq. (\ref{model}) are usually
complemented by the transition to matter-domination occurring at
$\eta_{dec}$. For $\eta >\eta_{dec}$ the scale factor evolves
parabolically (i.e. $a_{m}(\eta) \sim \eta^2$) in the matter epoch.
Our main goal in this paper will be to study the gravitational wave
spectra in these models and to study which kind of constraints arise
by
changing $H_1$ and $\gamma$. Notice that the time of the
inflation-stiff phase transition is $\eta_{1}= [a_i(t_{1})
H_i(t_1)]^{-1}$.

\renewcommand{\theequation}{3.\arabic{equation}}
\setcounter{equation}{0}
\section{Gravitational wave spectra from stiff phases}

The evolution of the scalar, vector and tensor fluctuations of a
given
background geometry can be directly discussed by perturbing (to
second
order in the amplitude of the fluctuations) the Einstein-Hilbert
action. An important property of the metric perturbations in FRW
backgrounds of the type defined in Eq. (\ref{line}) is that scalar,
vector and tensor modes are decoupled  \cite{bardeen}
(this feature does not hold in
the case of anisotropic background geometries \cite{mm}). This fact
means that, by defining the  fluctuations of the background metric
$\overline{g}_{\mu\nu}$ as
\begin{equation}
g_{\mu\nu}(\vec{x},\eta) = \overline{g}_{\mu\nu}(\eta) +
\delta g_{\mu\nu}(\vec{x},\eta),
\end{equation}
the metric fluctuation can be formally written as
\begin{equation}
\delta g_{\mu\nu} (\vec{x},\eta) = \delta g^{(S)}_{\mu\nu}
(\vec{x},\eta) + \delta g^{(V)}_{\mu\nu} (\vec{x},\eta)
+\delta g^{(T)}_{\mu\nu} (\vec{x},\eta),
\end{equation}
where $S,~V,~T$ stand respectively for, scalar, vector and tensor
modes. This classification refers to the way in which the fields
from
which $\delta g_{\mu\nu}(\vec{x},\eta)$ are constructed, change under
three
dimensional (spatial)
coordinate transformations on the constant (conformal)-time
hypersurface. Since
$\delta g_{\mu\nu}(\vec{x}, \eta)$
(being a  symmetric four-dimensional tensor of
rank 2) has ten independent component the scalar, vector and tensor
modes will be parametrized by ten independent space-time
functions. More specifically, scalar perturbations will be
parametrized by four independent scalar functions, vector
perturbations by two divergence-less three-dimensional vectors
(equivalent to four independent functions). Pure tensor modes of the
metric (corresponding to physical gravitational waves propagating in
a
homogeneous and isotropic background) can be constructed using a
symmetric three tensor $h_{ij}$ satisfying the constraints
\begin{equation}
h_{0\mu}=0,~~~h_{i}^{i}=0,~~~\nabla^{j} h_{ij} =0
\label{TT}
\end{equation}
( $\nabla_{i}$ denotes the covariant derivative with respect to
the three-dimensional metric). Notice that, with our conventions,
$\delta g^{(T)}_{\mu\nu} = h_{\mu\nu}$ and $ \delta g_{(T)}^{\mu\nu}
= -
h^{\mu\nu}$. The line element perturbed by the tensor modes can then
be written as
\begin{equation}
ds^2 = a^2(\eta) \biggl[ d\eta^2 - (\gamma_{ij} + h_{ij})
dx^{i}dx^j\biggr]
\end{equation}
(where $\gamma_{ij}$ is the spatial background metric).

Summarizing, we have  four functions for scalars, four
functions for vectors and two functions (the two polarizations of
$h_{ij}$) for the tensors. In total there are  ten independent
degrees of freedom describing the fluctuations
 as required by the tensor properties of the original
(unperturbed) metric. Moreover the conditions
expressed in Eq. (\ref{TT}) imply that the two physical
(independent) polarizations of $h_{ij}$
do not contain any pieces which transform as scalars or vectors
under three-dimensional rotations.
As a consequence of its definition, $h_{ij}$ is directly invariant
under infinitesimal coordinate transformations preserving the
tensorial character of the fluctuations \cite{grishchuk2,bardeen}.

In order to obtain the evolution equation of the metric fluctuations
there are at least two different
procedures. First of all one could think to perturb (to first order
in
the metric fluctuations) the Einstein equations. On the other hand
one
could also perturb the Einstein-Hilbert action
\begin{equation}
S = - \frac{1}{6 l^2_{P}} \int d^4x \sqrt{-g} R, ~~~R=
g^{\alpha\beta} R_{\alpha\beta},~~~g = {\rm det}
\bigg[g_{\mu\nu}\biggr],~~~l_{P}= M_{P}^{-1},
\label{action}
\end{equation}
to second order in the amplitude of the metric fluctuations
\cite{parker2}.

It is convenient to notice that the two procedure are certainly
equivalent but the perturbation of the action provides more
informations since it allows to isolate (up to total derivative
terms)
the normal modes of oscillation of the system which one might want to
normalize to the value of the quantum mechanical fluctuations.
By perturbing the action given in Eq. (\ref{action}) to second order
in the amplitude of the tensor fluctuations we obtain that, up to
total derivatives   \cite{mm},
\begin{equation}
\delta^{(2)} S^{(T)} = \frac{1}{24 l^2_{P}}\int d^4x
\sqrt{-\overline{g}} \biggl[
\partial_{\alpha}h_{ij}\partial_{\beta}h^{ij}
\overline{g}^{\alpha\beta} \biggr]
\label{second}
\end{equation}
(we remind  that in this and in the following formulas the shift
from upper to lower spatial indices [and viceversa] is done by using
the spatial background metric $\gamma_{ij}$ and its inverse).

For a wave moving in the $x^3=z$ direction in our coordinates one has
$h_{\oplus}=h_{1}^{1}= -h_{2}^{2}$ and $h_{\otimes} = h_{1}^{2} =
h_{2}^{1}$ and the perturbed action becomes:
\begin{equation}
\delta^{(2)} S^{(T)}=\frac{1}{12 l^2_{P}}\int d^4x
\sqrt{-\overline{g}}\biggl[
\partial_{\alpha}h_{\oplus}\partial_{\beta}h_{\oplus}
\overline{g}^{\alpha\beta}+\partial_{\alpha}h_{\otimes}\partial_{\beta}h_{\otimes} \overline{g}^{\alpha\beta} \biggr].
\label{second2}
\end{equation}
By now varying the perturbed action we get the evolution equation for
each polarization:
\begin{equation}
{\ddot{h}}_{\oplus} + 3 H {\dot{h}}_{\oplus} - \nabla^2
h_{\oplus}=0,~~~
{\ddot{h}}_{\otimes} + 3 H {\dot{h}}_{\otimes} - \nabla^2
h_{\otimes}=0.
\label{pol}
\end{equation}
{}From Eq. (\ref{second2}) it is also possible to deduce the form of
the
canonical normal modes, namely those modes whose Lagrangian reduces
to
the Lagrangian of two minimally coupled scalar fields (in flat space)
with time
dependent mass terms. Defining
\begin{equation}
\mu_{\oplus} = \frac{a h_{\oplus}}{\sqrt{6 l_{P}}},~~~
\mu_{\otimes} = \frac{a h_{\otimes}}{\sqrt{6 l_{P}}},
\end{equation}
 we can write, from Eq. (\ref{second2}), the action for the canonical
 normal modes
\begin{equation}
\delta^{(2)} S^{(T)} =\int d^3x~d\eta ~L,
\end{equation}
where (always up to total derivatives)
\begin{equation}
L= \frac{1}{2}\Biggl[\eta^{\alpha\beta}\partial_{\alpha}\mu_{\oplus}
\partial_{\beta}\mu_{\oplus}+
\eta^{\alpha\beta}\partial_{\alpha}\mu_{\otimes}
\partial_{\beta}\mu_{\otimes} + \biggl({\cal H}^2 + {\cal
H}'\biggr)\biggl(
{\mu_{\oplus}}^2 + {\mu_{\otimes}}^2\biggr) \Biggr]
\label{actionnormal}
\end{equation}
($\eta_{\alpha\beta}= {\rm diag}(1, -1, -1, -1)$ is the flat
space-time metric and ${\cal H} = [\log{a}]'$).
{}From Eq. (\ref{actionnormal}) we can derive the evolution equations
of
the canonical normal modes by taking the functional variation with
respect to $\mu_{\otimes}$ and $\mu_{\oplus}$ with  the result that:
\begin{equation}
\mu_{\otimes}''- \nabla^2 \mu_{\otimes} - \frac{a''}{a}
\mu_{\otimes}=0,~~~~
\mu_{\oplus}''- \nabla^2 \mu_{\oplus} - \frac{a''}{a} \mu_{\oplus}=0.
\label{eqcan}
\end{equation}
The normal modes obtained in Eq. (\ref{normal}) can be now
canonically
quantized. First of all we define the canonical momenta
\begin{equation}
\pi_{\oplus} = \frac{\partial L }{\partial \mu'_{\oplus}},~~~
\pi_{\otimes} = \frac{\partial L }{\partial \mu'_{\otimes}},
\end{equation}
leading to the Hamiltonian
\begin{equation}
Q=\int d^3x\biggl[ \pi_{\oplus} \mu'_{\oplus} +
\pi_{\otimes} \mu'_{\otimes} -L\biggl].
\label{ham}
\end{equation}
We then impose the (equal time) commutation relations between
the corresponding field operators:
\begin{equation}
\biggl[\hat{\mu}_{\oplus}(\vec{x},\eta),
\hat{\pi}_{\oplus}(\vec{y},\eta)\biggr]
= i\delta^{3}(\vec{x} -
\vec{y}),~~~\biggl[\hat{\mu}_{\otimes}(\vec{x},\eta),
\hat{\pi}_{\otimes}(\vec{y},\eta)\biggr]
= i\delta^{3}(\vec{x} - \vec{y}).
\label{comm}
\end{equation}
We want to remind that the way we got the Hamiltonian is quite naive.
In fact there is a (formally more correct) way of getting the
Hamiltonian for the perturbations \cite{ande}. One should in fact
start with the Hamiltonian in the superspace, fix the background, and
then obtain the Hamiltonian for the perturbations together with four
(independent) constraints (two for scalar perturbations and two for
vector perturbations). Therefore, in a fully consistent Hamiltonian
approach to perturbations there are no constraints arising from the
tensor modes.This
is the reason why our naive approach leads to the same result of the
Hamiltonian formalism for what concerns the tensor modes.
In the case od scalar fluctuations, however, our approach would
 be less correct since, in  the scalar case, the
 off-diagonal components of the
Einstein equations receive  contribution leading to
constraints. In order to get  the Hamiltonian for the scalar
modes (and the related algebra of the constraints) the (Hamiltonian)
 superspace approach is more compelling. We point out, in any case,
that also in our present approach it is possible to get the correct
result for the Hamiltonian  of the scalar modes by inserting, in the
perturbed action of the scalar fluctuations, the constraint equation
arising from the $(0i)$ components of the Einstein equations
\cite{muk}.

Promoting the classical Hamiltonian quantum mechanical operator
the evolution equations for the field operators become,
in the Heisenberg representation,
\begin{eqnarray}
&&i \hat{\mu}'_{\oplus} = \biggl[
\hat{\mu}_{\oplus},\hat{Q}\biggr],~~~
i \hat{\pi}'_{\oplus} = \biggl[ \hat{\pi}_{\oplus},\hat{Q}\biggr],
\nonumber\\
&&i \hat{\mu}'_{\otimes} = \biggl[
\hat{\mu}_{\otimes},\hat{Q}\biggr],~~~
i \hat{\pi}'_{\otimes} = \biggl[ \hat{\pi}_{\otimes},\hat{Q}\biggr].
\label{heis}
\end{eqnarray}
Notice that (as it has to be) the evolution equations
in the Heisenberg representation
are exactly identical (for the field operators) to the ones derived
for the classical fields in Eq. (\ref{eqcan}) once the explicit
expressions of the conjugated momenta (i.e. $\hat{\pi}_{\oplus} =
\hat{\mu}'_{\oplus}$ and $\hat{\pi}_{\otimes} =
\hat{\mu}'_{\otimes}$) are inserted back into Eq. (\ref{heis}).
We also point out that
the hermitian Hamiltonian of Eq. (\ref{ham}) is quadratic in the
field
operators and it belongs to a general class of time-dependent
Hamiltonians widely used in quantum optics \cite{gris4,gasg} in the
context of the parametric amplification of the vacuum fluctuations of
the electromagnetic field through  laser beams. The same
discussion we performed in the Heisenberg picture can be easily
translated to the Schroedinger picture where the final state of the
evolution is a many particle state unitarily connected to the initial
vacuum (a so called squeezed state \cite{gris4,gasg}).
Having fixed these standard notations we can expand the operators in
Fourier integrals
\begin{eqnarray}
&&\hat{\mu}_{\oplus} = \frac{1 }{(2\pi)^{3/2}}\int d^3 k
\biggl[ \mu_{\oplus} (k,\eta) \hat{a}_{\oplus}(\vec{k}) e^{i
\vec{k}\cdot\vec{x}} +
\mu^{\ast}_{\oplus}(k,\eta)\hat{a}^{\dagger}_{\oplus}(\vec{k})e^{-i
\vec{k}\cdot\vec{x}}\biggr],
\nonumber\\
&&\hat{\mu}_{\otimes} = \frac{1 }{(2\pi)^{3/2}}\int d^3 k
\biggl[ \mu_{\otimes} (k,\eta) \hat{a}_{\otimes}(\vec{k}) e^{i
\vec{k}\cdot\vec{x}} +
\mu^{\ast}_{\otimes}(k,\eta)\hat{a}^{\dagger}_{\otimes}(\vec{k})e^{-i
\vec{k}\cdot\vec{x}}\biggr]
\end{eqnarray}
(recall that, in our notations,
$\hat{h}_{\oplus,\otimes}(\vec{x},\eta =
\sqrt{6}[l_{P}/a] \hat{\mu}_{\oplus\otimes}(\vec{x},\eta)$).
{}From (\ref{comm}) $\hat{a}_{\oplus}$ and $\hat{a}_{\otimes}$ obey
the
following commutation relations
\begin{equation}
\biggl[\hat{a}_{\oplus}(\vec{k}),
\hat{a}^{\dagger}_{\oplus}(\vec{k}')\biggr]
=
\delta^{(3)}(\vec{k}-\vec{k}'),~~~\biggl[\hat{a}_{\otimes}(\vec{k}),
\hat{a}^{\dagger}_{\otimes}(\vec{k}')\biggr] =
\delta^{(3)}(\vec{k}-\vec{k}'),
{}~~~\biggl[\hat{a}_{\oplus}(\vec{k}),\hat{a}_{\otimes}(\vec{k}')\biggr]=0.
\end{equation}
Therefore the  evolution equation for the Fourier amplitudes
$\mu_{\oplus}(k,\eta)$ and $\mu_{\otimes}(k,\eta)$ become
\begin{equation}
\mu_{\oplus,\otimes}'' + \biggl[ k^2 -
f(\eta)\biggr]\mu_{\oplus,\otimes}
=0,~~~f(\eta)=\frac{a''}{a} \equiv a^2( \dot{H} + 2 H^2)
\label{normal}
\end{equation}
(where, for the Fourier amplitudes, we define $\mu(k,\eta) = a
h(k,\eta)$).

By taking the functional derivative of the action reported in Eq.
(\ref{second}) with respect to the
background metric we indeed find an effective energy-momentum tensor
of the  fluctuations which reads:
\begin{equation}
\tau_{\mu\nu} = \frac{1}{6~l^2_{P}}\Biggl[ \partial_{\mu}h_{\oplus}
\partial_{\nu}h_{\oplus} +
\partial_{\mu}h_{\otimes}\partial_{\nu}h_{\otimes} -
\frac{1}{2}\overline{g}_{\mu\nu}\biggl(\overline{g}^{\alpha\beta}
\partial_{\alpha} h_{\oplus}
\partial_{\beta}h_{\oplus} +\overline{g}^{\alpha\beta}
\partial_{\alpha} h_{\otimes}
\partial_{\beta}h_{\otimes}\biggl)\Biggr].
\end{equation}
If we define the vacuum state vector $|0_{\oplus}0_{\otimes}\rangle$
which is annihilated by $\hat{a}_{\oplus}$ and  $\hat{a}_{\otimes}$
we
obtain that the energy density of the produced gravitons will be
given
by
\begin{eqnarray}
&&\rho_{GW}(\eta) = \langle
0_{\otimes}0_{\oplus}|\tau_{oo}|0_{\oplus}0_{\otimes}\rangle=
\nonumber\\
&&\frac{1}{16\pi^3 ~a^2 } \int d^3 k \Biggl[ |h'_{\oplus}(k,\eta)|^2
+
|h'_{\otimes}(k,\eta)|^2 +
k^2 \biggl(|h_{\oplus}(k,\eta)|^2 + |h_{\otimes}(k,\eta)|^2
\biggr)\Biggr]
\label{aven}
\end{eqnarray}
In order to compute the GW spectra produced by the transition of the
background from an inflationary phase to a decelerated, stiff, phase
we have to solve the evolution equation (\ref{normal}).
{}From Eq. (\ref{normal}) we clearly see that the evolution equation
of
the (tensor) normal modes of the geometry is determined not only by
$H^2$ but also by $\dot{H}$. Therefore, depending upon the sign of
$\dot{H}$ different inflationary models
will give different (large scales) spectral distributions of the
amplified
gravitational waves.

In the three phases defined in Eq. (\ref{model}) the evolution
equation (\ref{normal}) reads :
\begin{eqnarray}
&&\mu'' + \biggl[ k^2 - \frac{2}{\eta^2}\biggr]\mu =0,~~~~~~~~\eta<
-\eta_{1},
\nonumber\\
&&\mu'' + \biggl[ k^2 - \frac{\beta(\beta -1)}{[\eta + (\beta +
1)\eta_1]^2}\biggr]\mu =0,~~~~- \eta_1< \eta < \eta_r,
\nonumber\\
&& \mu'' + k^2 \mu=0,~~~~~~~~~~~~~~~~~~~\eta_{dec} <\eta <\eta_r
\label{muequation}
\end{eqnarray}
(notice that we dropped the subscript
referring to each polarization).
\begin{figure}
\begin{center}
\begin{tabular}{|c|}
      \hline
      \hbox{\epsfxsize = 9.5 cm  \epsffile{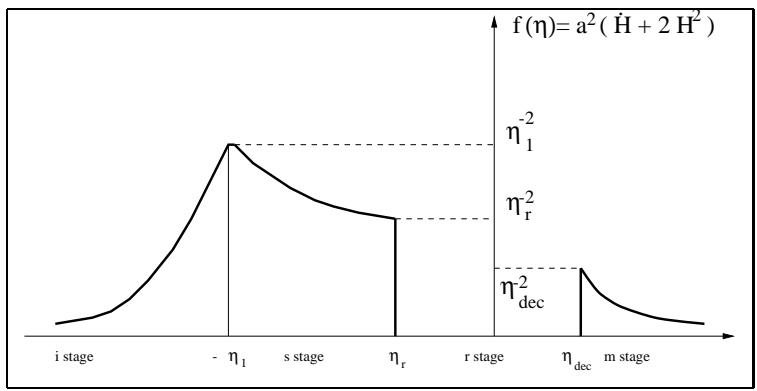}} \\
      \hline
\end{tabular}
\caption[a]{The ``effective potential'' $a''/a$ appearing in the
Schroedinger-like equation (\ref{normal}) is reported as a function
of
the conformal time coordinate. In the standard scenario, the
intermediate (stiff) phase is absent and, therefore, all the modes
$k$
will reenter during the radiation epoch leading (at high frequencies)
to the usual Harrison-Zeldovich (flat) energy spectrum. In our case
the modes re-entering during the radiation dominated era
($k<\eta^{-1}_{r}$) will always have flat energy spectrum. At the
same
time, some modes going under the potential barrier during the de
Sitter phase ($i$-stage) and re-entering the horizon during the stiff
phase ($s$-stage) will lead to a modification of the high frequency
branch of the spectrum which will become mildly increasing (see
Eqs. (\ref{spectra1})-(\ref{spectra2}) and Eq. (\ref{grow})). We
stress that at large ($10^{-18} {\rm Hz} < \omega < 10^{-16} {\rm
Hz}$) and intermediate ($10^{-16} {\rm Hz} < \omega <\omega_{r}$)
(present) physical frequencies,
the spectrum does not change with respect to the usual
results \cite{grishchuk1,all,sahni,grishchuk4}
since the corresponding modes will not ``feel'' the presence of
the stiff phase.}
\label{pot}
\end{center}
\end{figure}
The solution of the evolution equations in the three regions can be
written in terms of Hankel functions \cite{abr,bat}.
Since we are dealing with the
Fourier amplitudes of the normal modes of oscillation we normalize
them directly to the quantum
mechanical noise level ($\sim 1/\sqrt{k}$) for $\eta < - \eta_1$:
\begin{eqnarray}
&&\mu_{I}(u) = \frac{1}{\sqrt{2k}} \sqrt{u} H^{(2)}_{3/2}(u),
{}~~~~~\eta< -\eta_1,
\nonumber\\
&&\mu_{II}(v) = \frac{1}{\sqrt{2k}} \sqrt{v}\biggl[b_{+}
H^{(2)}_{\nu}(v) +
b_{-} H^{(1)}_{\nu}(v)\biggr]~~~~- \eta_1< \eta < \eta_r,
\nonumber\\
&&\mu_{III}(u) = \frac{1}{\sqrt{2k}}\biggl[ c_{+} e^{- iu} + c_{-}
e^{i
u}\biggr]~~~~~\eta_{dec} <\eta <\eta_r,
\label{solution}
\end{eqnarray}
where
\begin{equation}
2 \nu = |2 \beta -1|,~~~u=k\eta,~~~v=k[\eta + (\beta + 1)\eta_1].
\end{equation}
Notice that the Wronskian normalization of the solutions imposes that
$|b_{+}|^2 - |b_{-}|^2 =1$ and $|c_{+}|^2 - |c_{-}|^2 =1$.
See also Fig. \ref{pot}
for a pictorial description of the amplification process.
By computing the coefficients $b_{-}$ and $c_{-}$ we can have an
estimate of the energy spectrum of the produced gravitons
re-entering the horizon, respectively, after $\eta_{1}$ (i.e. right
after the
completion of the inflationary phase) and after $\eta_{r}$
(i.e. immediately after the beginning of the radiation dominated
epoch).
Notice that $\eta_{r}$ appearing in Eq. (\ref{solution}) is not
free parameter  since the
duration of the stiff phase ending in $\eta_{r}$ is determined
essentially by the backreaction of the modes re-entered right after
$\eta_{1}$. In order to further clarify this point we have to compute
the GW energy spectra produced right after $\eta_1$.
The energy density of the produced
gravitons per logarithmic interval of longitudinal momentum can be
obtained from Eq. (\ref{aven}) :
\begin{equation}
\rho(\omega,t)= \frac{d\rho_{GW}(\omega,t)}{d\log{\omega}} ,
\label{log}
\end{equation}
where $\omega = k/a$. Sometimes $\rho(\omega,t)$ is called
logarithmic energy spectrum. Notice that, one can define the energy
density
 of the gravitational wave background (in a semiclassical
description)
 through the energy momentum pseudo-tensor of the amplified tensor
 modes. In the
 approach of the energy momentum pseudo-tensor the quantum mechanical
expectation values are replaced by ensemble averages over a
 distribution of stochastic variables whose two-point function has to
 be specified separately by requiring that each Fourier amplitude of
 the tensor modes is independent on the others in momentum space.
This ``stochastic''  condition on the Fourier amplitudes is the
 result, in a fully quantum mechanical approach, of the unitarity of
 the graviton production process
 driven by the hermitian Hamiltonian (\ref{ham}). We stress that, if
 our initial state is not the vacuum (but for example a  non-pure
 state characterised by a specific density matrix \cite{gasg}),
the expression given in
 Eq. (\ref{aven}) for the averaged energy density will also be
different.

{}From Eq. (\ref{aven}) and using the definition (\ref{log}) we have
that in each phase of the model the energy spectra can be written as:
\begin{eqnarray}
&&\rho(\omega,t)\simeq \frac{1}{2\pi^2}H^4_{1}
\biggl(\frac{\omega}{\omega_{1}}\biggr)^{4}|b_{-}|^2
\biggl[\frac{a_{1}}{a}\biggr]^4
,~~~\eta>-\eta_1,
\nonumber\\
&&\rho(\omega,t) \simeq  \frac{1}{2\pi^2} H^4_{1}
\biggl(\frac{\omega}{\omega_{1}}\biggr)^{4} |c_{-}|^2
\biggl[\frac{a_{1}}{a}\biggr]^4
{}~~~\eta>\eta_r.
\label{energy}
\end{eqnarray}
In order to compute the coefficients $|b_{-}|$ and $|c_{-}|$ we use
the sudden approximation, namely we will mainly consider the
amplification of the modes $k\eta_1 \laq 1$ \cite{verda}. A
 possible technique is
to match, in $\eta_{1}$ and $\eta_{r}$ the exact solutions given by
Eq. (\ref{solution}) and their first derivatives. Therefore, imposing
the conditions
\begin{eqnarray}
&&\mu_{I}(u_1) = \mu_{II}(v_1),~~~\mu'_{I}(u_1) =
\mu'_{II}(v_1),
\nonumber\\
&&\mu_{II}(v_r) = \mu_{III}(u_{r}),~~~\mu'_{II}(v_r) =
\mu'_{III}(u_{r}),
\end{eqnarray}
we obtain an expression for the amplification coefficients which we
evaluate in the sudden approximation:
\begin{eqnarray}
|b_{-}| &\simeq & \biggl(\frac{\omega}{\omega_{1}}\biggr)^{-
\frac{3}{2} -
\nu},~~~~~~~~~~~~~\omega_{r}< \omega<\omega_1,
\nonumber\\
|c_{-}| &\simeq &
\biggl(\frac{\omega}{\omega_{r}}\biggr)^{-\frac{1}{2}}
\biggl(\frac{\omega}{\omega_{1}}\biggr)^{- \frac{3}{2}}
\biggl(\frac{\omega_{1}}{\omega_{r}}\biggr)^{\nu},~~\omega_{dec}<
\omega<\omega_r,
\label{coefficient1}
\end{eqnarray}
(notice that $u_{1,r}=k\eta_{1,r}$ and $v_{1,r} = k[\eta_{1,r} +
(\beta+1)\eta_1]$).
These expression are correct as long as the effective equation of
state parameterizing the evolution of the sources in the intermediate
phase has $\gamma<1$. For $\gamma=1$ (corresponding to $\beta =
\frac{1}{2}$, $\alpha= \frac{1}{3}$ and $\nu =0$),
 the solution of Eq. (\ref{muequation})
in the intermediate phase (i.e. what we call $\mu_{II}(v)$ in
Eq. (\ref{solution})) is given in terms of $H^{(1,2)}_{0}(v)$, and,
therefore, the amplification coefficients get logarithmically
corrected (in the sudden approximation), and the final result is :
\begin{eqnarray}
|b_{-}| &\simeq & \biggl(\frac{\omega}{\omega_{1}}\biggr)^{-
\frac{3}{2}}\log{\biggl(\frac{\omega_{1}}{\omega_{r}} \biggr)}
,~~~\omega_{r}< \omega<\omega_1,
\nonumber\\
|c_{-}| &\simeq &
\biggl(\frac{\omega}{\omega_{r}}\biggr)^{-\frac{1}{2}}
\biggl(\frac{\omega}{\omega_{1}}\biggr)^{- \frac{3}{2}}
\log{\biggl(\frac{\omega_{1}}{\omega_{r}} \biggr)}
\label{coefficient2},~~~\omega_{dec}< \omega<\omega_r.
\end{eqnarray}
Notice that $\omega_{dec}\simeq  10^{-16}~{\rm Hz}$
 is the frequency corresponding to
$\eta_{dec}$, namely to the moment of the transition to the
matter-dominated epoch. Of course the gravitational wave spectra will
 have a further (fourth) branch corresponding to the modes
re-entering
 the horizon during the radiation dominated epoch \cite{review}.
This infrared
 branch of the spectrum concerns frequencies $\omega_{0}<\omega
 <\omega_{dec}$ where $\omega_0 \simeq 10^{-18}~{\rm Hz}$ is the
 frequency corresponding to the present horizon. It is well known
that
 in the infra-red branch the spectrum turns out to be steeply
 decreasing as $\omega^{-2}$ \cite{grishchuk4,review}.
Since the purpose of our investigation are
 the effects of the produced gravitons during the stiff phase we will
 neglect, in the following considerations, the infrared branch.

By inserting the obtained expressions of the amplification
coefficients
in the expression for the energy spectrum of Eq. (\ref{energy}) we
get
the main result of this Section
\begin{eqnarray}
&&\rho(\omega,t) \sim \frac{1}{2\pi^2} H^4_{1}
\biggl(\frac{\omega}{\omega_{1}}\biggr)^{ 1 - 2 \nu}
\Biggl[\frac{a_1}{a}\Biggr]^{4},~~~\omega_{r}< \omega<\omega_1,
\nonumber\\
&&\rho(\omega,t) \sim \frac{1}{2\pi^2} H^4_{1}
\biggl(\frac{\omega_r}{\omega_{1}}\biggr)^{1 - 2 \nu}
\Biggl[\frac{a_1}{a}\Biggr]^{4},~~~\omega_{dec}< \omega<\omega_r.
\label{spectra1}
\end{eqnarray}
Again, this expression is logarithmically corrected in the
 $\gamma\rightarrow 1$ limit:
\begin{eqnarray}
&&\rho(\omega,t) \sim \frac{1}{2\pi^2} H^4_{1}
\biggl(\frac{\omega}{\omega_{1}}\biggr)
\log^2{\biggl(\frac{\omega}{\omega_{1}}\biggr) }
\Biggl[\frac{a_1}{a}\Biggr]^{4},~~~\omega_{r}< \omega<\omega_1,
\nonumber\\
&&\rho(\omega,t) \sim \frac{1}{2\pi^2} H^4_{1}
\biggl(\frac{\omega_r}{\omega_{1}}\biggr)
\log^2{\biggl(\frac{\omega_r}{\omega_{1}}\biggr) }
\biggl[\frac{a_1}{a}\biggr]^{4},~~~\omega_{dec}< \omega<\omega_r.
\label{spectra2}
\end{eqnarray}
Concerning Eqs. (\ref{spectra1}) and (\ref{spectra2}) few comments
are
in order. First of all we can notice that the
energy-spectra are scale-invariant for $\omega<\omega_{r}$ This
branch
of the spectrum corresponds to modes which went out of the horizon
during the initial de Sitter phase and re-entered during the
radiation
dominated phase. There is, therefore, no surprise for this behaviour
which does correspond to  the usual Harrison-Zeldovich spectrum of
the
stochastic GW backgrounds. Moreover, we point out that for
$\omega_r\rightarrow \omega_1$ (i.e. point-like stiff phase) we
reproduce
the usual results well known in the context of the theory of
graviton production when a de Sitter phase is suddenly followed by a
radiation dominated phase \cite{grishchuk1,grishchuk4,review}.
For the modes leaving the horizon during the de Sitter phase and
re-entering  during the stiff phase  ($\omega_r <\omega <\omega_1$)
we get the curious result that the energy spectrum increases as a
function of the  frequency. This feature of our result can be easily
understood by bearing in mind that the ultraviolet branch of the
spectrum goes as
\begin{equation}
\biggl(\frac{\omega}{\omega_1}\biggr)^{1- 2\nu},~~~
\nu = \frac{3(1 -\gamma)}{2(3\gamma+1)}.
\label{grow}
\end{equation}
If $\frac{1}{3}<\gamma\leq 1$ we have that $2\nu<1$ and the spectrum
 is always mildly increasing with maximal
slope $(\omega/\omega_1)$ (for $\gamma=1$). The minimal slope
 corresponds to the case $\gamma=1/3$ (flat case) where the flat
 spectrum is recovered for all the (present) frequency range.
This peculiar behaviour is indeed not so
strange. This peculiar feature of stiff models was indeed noticed
long
ago \cite{grishchuk3}. In our case the only difference is that prior
to the stiff phase there is a de Sitter phase and, therefore, the
calculation of the amplification coefficients involves a further
transition. This further transition modifies the high energy
behaviour
of the spectrum which still increases but more mildly if compared to
the case where the de Sitter phase was absent \cite{grishchuk3}.
If the inflationary phase is not de
Sitter but power-law (i.e. $\dot{H} <0$) or superinflationary
\cite{max2}
 ($\dot{H} >0$) the spectra of the modes re-entering in the
radiation epoch will be crucially modified and their  amplitude will
be
always subjected to the large scale constraint ($H_1\laq 10^{-6}
M_{P}$) provided the inflationary phase is either de Sitter-like or
power law. In the case of superinflation (leading to increasing
energy
 spectra), the nucleosynthesis
constraint will always be the most stringent one \cite{review,BBN}
since it
involves the integrated energy density and it applies at all the
frequencies.

The background energy decreases, during the stiff phase, as
$H^2_1 M^2_{P}(a_1/a)^{3(\gamma+1)}$, whereas the radiation stored in
GW decreases like $(a_1/a)^4$ and therefore, at
some stage, the graviton radiation will become dominant.
To compute precisely this moment we have to integrate the
graviton spectrum over all the modes, insert it back in the Einstein
equations and, finally solve the modified Einstein equations.
By defining $\epsilon = H_1/M_P$, a
 simple argument based on the calculations reported in the previous
Section shows that, since the energy spectra are increasing, the most
significant contribution of the hard gravitons to the energy density
(integrated over the whole spectrum) will occur for
$\omega\sim \omega_1$. This energy density will turn the stiff
background into radiation at a critical value of the scale factor
$a_{r}
\sim  \epsilon^{2/(1- 3 \gamma)} a_{1}$ and
the stiff fluid will correspondingly turn radiation
(i.e. $\gamma\rightarrow 1/3$). Taking now into account that during
the stiff phase $a(t) \sim t^{2/[3(\gamma+1)]}$ we have that,
according
to our estimate the back reaction effects will become significant at
a
curvature scale $H_{r} \sim \epsilon^{(6\gamma+2)/(3\gamma-1)}
M_{P}$. For example, if we take $\gamma =1$,
 the
curvature scale at which the transition to radiation takes place is
$H_{r} \sim \epsilon^4 M_{P}$. In frequency, the
length of the stiff phase (with $\gamma=1$) will then be
$\omega_{r}/\omega_{1} \sim \eta_1 a_1/\eta_r a_r \sim
 \epsilon^3$. Thus, if one wants to reach
(within a stiff phase with $\gamma=1$) the hadronic curvature scale
$H_{had} \sim 10^{-40} M_{P}$ we should fine-tune $\epsilon \sim
10^{-10}$.

The maximal scale $H_1$ is certainly constrained. If a de Sitter
phase is immediately followed by a radiation dominated phase, the GW
energy
spectrum  is flat for
$\omega_{dec}<\omega<\omega_{1}$ (notice that, today, $\omega_{dec}
\sim
10^{-16}~{\rm Hz}$ and  $\omega_{1}
\sim 10^{11}\sqrt{\epsilon}~{\rm Hz}$). For $\omega_0 <\omega
<\omega_{dec}$ the spectrum decreases and, therefore, $\epsilon \laq
10^{-6}$ which is exactly the bound reported in Eq. (\ref{bound})
coming from the tensor contribution to the CMBR anisotropy.

If the de Sitter phase is followed by a stiff phase the spectra are
growing for $\omega_r <\omega<\omega_{1}$. Take,  for instance, the
case $\gamma=1$ where $\rho(\omega,t) \sim H^4_1(\omega/\omega_1)$
up to logarithmic corrections. In this branch, since the spectrum
increases, the most significant constraint comes from the bound
energy
density in relativistic degrees of freedom at the nucleosynthesis
epoch \cite{review,BBN}. Thus the bounds on $\epsilon$
might be a bit different (even if not by much due to the very mild
increase of the spectral energy density). Today, in the range
$\omega_{dec} <\omega <\omega_{r}$ the spectral energy
density (in critical units) would be
  $\Omega_{GW}(\omega,t)\sim ~10^{-4}\epsilon^2
(\omega_{r}/\omega_1)$. Now the large
scale bound imposes that, in this phase, $\Omega_{GW}\laq
10^{-13}$. Taking now into account that $\omega_{r}/\omega_1 \sim
\epsilon^3$ (for the case $\gamma=1$) we have that
$\Omega_{GW}(\omega,t) \sim \epsilon^{5}$. Imposing now $\epsilon
\laq ~10^{-3}$ we get that, for $\omega_{dec}<\omega<\omega_{r}$,
$\Omega_{GW}(\omega,t) < 10^{-14}$. The argument we just discussed
can be easily extended to the entire class of stiff models
(i.e. $1/3<\gamma <1$).

In conclusion
the possibility of having a stiff, post-inflationary phase implies
necessarily a `'reheating'' driven by gravitational waves re-entering
the horizon during the stiff phase. In order to have a significantly
long stiff phase, however, fine-tuning is strictly required making
these models perhaps less attractive. We will elaborate on this point
in the next Section.

\renewcommand{\theequation}{4.\arabic{equation}}
\setcounter{equation}{0}
\section{Back reaction effects}

In order to estimate the length of the transition between the stiff
phase and the radiation dominated phase induced by the hard
gravitons,
we re-write Eq. (\ref{normal}) in a slightly
different form, namely \cite{parker}
\begin{equation}
\biggl[\frac{d^2 }{d \tau^2} +
\Omega_{k}^2(\tau)\biggr]h_{\oplus,\otimes}=0
,~~~d\eta=\frac{dt}{a} = a^2 d\tau,
\label{neweq}
\end{equation}
where $\Omega_k(\tau) = \sqrt{-g} \omega = k a^2$ and
$h= a^{-1} \mu $.
A formal solution to this equation can be written as
\begin{equation}
h(k,\tau) = \frac{1}{\sqrt{2 \Omega_{k}}} \biggl[C_{+}h_{-}(k,\tau) +
C_{-} h_{+}(k,\tau)\biggr],
\label{formal}
\end{equation}
with
\begin{equation}
h_{\pm}(k,\tau) = \exp{\biggl[\pm i \int \Omega_{k} d\tau\biggr]},
\end{equation}
(notice that $C_{+}$ and $C_{-}$ are complex functions of $\tau$;
thanks to the Wronskian normalization condition we also have that
$|C_{+}|^2 - |C_{-}|^2=1$).
By inserting the solution given in Eq. (\ref{formal}) back into
Eq. (\ref{neweq}) we can obtain an evolution equation for the two,
time-dependent coefficients, $C_{\pm}$ \cite{zeldovich2}
\begin{equation}
\frac{d C_{+}}{d\tau}= \frac{1}{2} \frac{d\log{\Omega_{k}}}{d\tau}
h^2_{+} C_{-},~~~~~~
\frac{d C_{-}}{d\tau}= \frac{1}{2} \frac{d\log{\Omega_{k}}}{d\tau}
h^2_{-} C _{+},
\label{coeff}
\end{equation}
(from now on we will drop the subscript referring to the two
polarizations).
We are now going to solve these equations for the modes $k\eta_1 \laq
1$.
It is easy to show that this sudden approximation  used in the
previous section corresponds, in the language of Eq. (\ref{neweq}) to
the small $\int^{\tau_{1}} d \tau \Omega_{k}(\tau)$ limit. In fact
\begin{equation}
\int^{\tau_{1}} \Omega_{k}(\tau) d\tau = \int^{\tau_1} k a^2
d\tau\equiv \int^{\eta_{1}} k d \eta \sim k\eta_1
\end{equation}
(the last two equalities follow from the definition of $d{\tau}$ in
terms of the conformal time coordinate : $d\tau = d\eta/a^2$).

In this approximation we can expand the $h_{\pm}$ appearing in
Eq. (\ref{coeff}) and we find, to first order in $\int d\tau
\Omega_{k}$,
\begin{equation}
\frac{d C_{+}}{d\tau}= \frac{1}{2} \frac{d\log{\Omega_{k}}}{d\tau}
\biggl[ 1 + 2 i \int \Omega_{k} d\tau\biggr] C_{-},~~~~
\frac{d C_{-}}{d\tau}= \frac{1}{2} \frac{d\log{\Omega_{k}}}{d\tau}
\biggl[ 1 - 2 i \int \Omega_{k} d\tau\biggr] C _{+}.
\label{coeff2}
\end{equation}
By now linearly combining two previous equations we find:
\begin{eqnarray}
&&\frac{d}{d\tau}\biggl[C_{+} + C_{-}\biggr] = \frac{1}{2}
\frac{d\log{\Omega_{k}}}{d\tau}
\biggl[C_{+} + C_{-}\biggr]+...
\nonumber\\
&&\frac{d}{d\tau}\biggl[C_{+} - C_{-}\biggr] = -\frac{1}{2}
\frac{d\log{\Omega_{k}}}{d\tau}
\biggl[C_{+} - C_{-}\biggr]+ ...
\label{coeff3}
\end{eqnarray}
In Eq. (\ref{coeff3}) the ellipses stand for other terms which are of
higher order in $\int^{\tau_1} \Omega_{k}(\tau)d\tau\sim k\eta_1$ and
which are negligible in the sudden approximation.
The solution to Eq. (\ref{coeff3}) can be easily found in terms of
two
(arbitrary) complex coefficients
\begin{equation}
C_{+} + C_{-} = 2 Q_1 \sqrt{\Omega_{k}},~~~C_{+} -  C_{-} =
2 Q_2 \frac{1}{\sqrt{\Omega_{k}}}.
\label{firstorder}
\end{equation}
As we previously mentioned, the Wronskian condition imposes that
$|C_{+}|^2 - |C_{-}|^2 =1$. Now this last condition has to hold order
by order in $k\eta_1$ and, therefore, inserting the $C_{\pm}$ of
Eq. (\ref{firstorder}) into $|C_{+}|^2 - |C_{-}|^2 =1$ we obtain a
condition on $Q_{1}$ and $Q_{2}$ valid to first oder in $k\eta_1$:
\begin{equation}
C_{+} = \biggl[ Q_1 \sqrt{\Omega_{k}} + \frac{Q_2}{ \sqrt{\Omega_{k}}
}\biggr]
,~~~C_{-} = \biggl[ Q_1 \sqrt{\Omega_{k}} -
\frac{Q_2}{\sqrt{\Omega_{k}}}\biggr],~~~Q^{\ast}_{1} Q_2 + Q_1
Q^{\ast}_2 = \frac{1}{2}.
\label{fin}
\end{equation}
Since the amount of amplification of the GW is essentially
determined,
in this approach, by $|C_{-}|^2$ we fix $Q_1$ and $Q_{2}$ by
requiring
that at the time $\eta_{1}$ (or $t_1$) $|C_{-}(k,t_1)| =0$. Thus,
from
Eq. (\ref{fin}) we  get that
\begin{equation}
|Q_{1}| = \frac{1}{2 \sqrt{\Omega_k(t_{1})}},~~~|Q_{2}| =
\frac{\sqrt{\Omega_{k}(t_{1})}}{2},~~~\Omega_{k}(t_1) = a^2_1 k.
\end{equation}
Using Eq. (\ref{formal}) and summing the polarizations we get that
\begin{equation}
\rho_{GW}(t)=  \frac{1}{8~\pi^3}\int d^3 \omega \omega\biggl[
|C_{+}(\omega,t)|^2 + |C_{-}(\omega,t)|^2\biggr]=
\frac{1}{8~\pi^3}\int \omega d^3 \omega\biggl[2 |C_{-}(\omega,t)|^2 +
1\biggr]
\label{en2}
\end{equation}
(where in the last equality we used the Wronskian normalization
condition).
To the lowest order in $k\eta_1$, using Eqs. (\ref{fin}) and
(\ref{firstorder}), and
expressing all the quantities in terms of the corresponding physical
momenta $\omega$, we get
\begin{equation}
\rho_{GW}(t) = \frac{1}{16~\pi^3}\int \omega d^3 \omega\biggl[
\frac{\Omega_{\omega}(t)}{\Omega_{\omega}(t_1)} +
\frac{\Omega_{\omega}(t_1)}{\Omega_{\omega}(t)} \biggr].
\label{fin2}
\end{equation}

Now we want to compute how much energy is present at a generic time
$t$ inside the horizon during the stiff phase. First of all we will
outline the formal solution of the problem and secondly we will do an
explicit calculation.

The total energy density in GW at a generic time $t$ will be given by
the explicit momentum integral indicated in Eq. (\ref{fin2}). This is
however not the end of the story. At any given time different GW will
re-enter during the stiff phase and therefore the total energy stored
in GW which should appear at the right hand side of the Einstein
equations is the sum over all waves re-entering at different times
after $t_1$. This means that the quantity we  we should compute is
\cite{parker} the total energy density of the gravitational waves
re-entered
at some generic time $t$ during the stiff phase. We then
have  that the total energy density of the
graviton background is given by the modes re-entering at $t$ summed
to
the (red-shifted) energy density of those modes which re-entered
during the period $t_1 <t'<t$:
\begin{equation}
\rho_{{\rm tot}}(t) = - \int_{t_{1}}^{t}
\biggl[\frac{a(t')}{a(t)}\biggr]^4
\biggl[ \frac{\partial \rho_{GW}(\omega_{m}(t'),t')}{\partial
\omega_{m}}\bigg]_{a} \frac{\partial \omega_{m}}{\partial t'} d t'
\label{integro}
\end{equation}
Concerning Eq. (\ref{integro}) few comments are in order. First of
all
$\rho_{GW}(\omega_m,t)$ denotes the energy density in GW (see
Eq. (\ref{fin2})) integrated until a ``running'' ultra-violet cut-off
$\omega_{m}(t) \sim \frac{1}{t}\sim H(t)$. The partial derivatives
appearing in Eq. (\ref{integro}) simply reflect the fact that we are
summing up the energy of GW re-entering the horizon at different
moments $t>t_{1}$. In order to do this we have exactly to slice the
horizon in many infinitesimal portions, and compute, for each
portion,
the corresponding increment in the GW energy. In computing the
increment in the GW energy with respect to the cut-off  we keep the
scale factor constant (and this is the reason of the subscript
appearing in Eq. (\ref{integro}).
We want now to compute the frequency integral of Eq. (\ref{fin2}).
As it is well known this integral is divergent in the limit of large
frequencies. This is simply a consequence of the fact that the vacuum
leads to an infinite energy density which has to be properly
subtracted. The spectrum of the vacuum fluctuations can be simply
obtained by putting $|C_{-}(\omega,t_{1})|=0$ in Eq. (\ref{en2}).
This limit
corresponds to the absence of amplification, and therefore the only
fluctuations contributing to the energy density are the ones
associated with the vacuum modes with logarithmic energy spectrum
proportional to $\omega^4$.
 In principle, in our case we have a (physical) ultraviolet
cut-off in the spectrum provided by the scale where inflation
stops. This maximal frequency is $k \simeq {\eta_{1}}^{-1}$. We then
expect our results to be insensitive to the particular
renormalization scheme. Since however we want to have (in
Eq. (\ref{integro})) the possibility of a cut-off running with time
we
will examine briefly this issue which was actually investigated in
the
past, for homogeneous
cosmological backgrounds, within (at least)  two different
approaches.
In Ref. \cite{zeldovich2} this problem was tackled using a
regularization scheme strongly reminiscent of the Pauli-Villars
method. In Ref. \cite{fulling} the same problem was discussed within
the so called adiabatic regularization scheme (see also
\cite{parker}).
The two methods were shown to produce equivalent results
\cite{fulling}. The purpose of this paper is not to check which
regularization scheme is better to use in curved space. Our approach
is more pragmatic: we want to get an estimate not only of the maximal
duration of the stiff phase but also of the transition time between
the stiff and the radiation dominated epoch.
 We want to know how long will it take for the
stiff fluid to turn into radiation. In this spirit we will firstly of
all use the adiabatic regularization scheme (without including higher
order subtractions involving double time derivatives of the Hubble
parameter). Secondly we will re-compute our back reaction effects on
the background evolution
without including the subtractions and using a (naive) cut-off
regularization.

In the adiabatic scheme the regularized energy density reads
\begin{equation}
\overline{\rho}_{GW}(t) = \frac{1}{8~\pi^3}\int \omega d^3
\omega\biggl[ F_1(\omega,t) - F_2(\omega,t)\biggr]
\label{enreg}
\end{equation}
where
\begin{equation}
F_1(\omega,t)
=\frac{1}{2}\biggl[\biggl(\frac{\Omega_{\omega}(t)}{\Omega_{\omega}(t_{1})
}\biggr) +
\biggl(\frac{\Omega_{\omega}(t_1)}{\Omega_{\omega}(t)}\biggr)
\biggr],
{}~~~F_2(\omega,t) = \biggl[1 + \frac{1}{2}
\biggl(\frac{H}{\omega}\biggr)^2 +
\frac{1}{8} \frac{\Sigma(t)}{\omega^4}\biggr]
\end{equation}
We notice that $F_1(\omega,t)$ is simply the non-regularized
contribution
and it is exactly equal to the one we computed Eq. (\ref{en2}). On
the other hand, $F_2(\omega,t)$ comes from the adiabatic
regularization
 \cite{fulling} and it does contain the divergent contribution which
we want to subtract. Notice that, in the expansion there are terms
with
higher derivatives  since
\begin{equation}
\Sigma(t) = \biggl[ {\dot{H}}^2 + 2 H^4 - 2 \ddot{H} H - 4 H^2
\dot{H}\biggr]
\end{equation}
Now, our underlying theory is the Einstein theory (with only linear
curvature terms in the action (\ref{action})). Consequently, it is
not compatible to include
subtractions involving four derivatives.  Therefore we will not
include them in the subtraction.
We can integrate the energy density keeping a running  ultra-violet
cut-off $\omega_{m} \sim H(t)\laq H_{t_1}$
\begin{equation}
\overline{\rho}_{GW}(t) = \frac{1}{4~\pi^2} \int^{\omega_{m}(t)}
\omega^4
 \Biggl\{ \biggl[ \biggl( \frac{a}{a_1}\biggr)-
\biggl(\frac{a_1}{a} \biggr)^2 \biggr]^2
-\frac{H^2}{\omega^2}\Biggr\}
d\log{\omega},
\end{equation}
with the result that
\begin{equation}
\overline{\rho}_{GW}(t) = \frac{\omega^4_{m}}{16 \pi^2} \Biggl\{
\bigg[\frac{a}{a_{1}} - \frac{a_1}{a}\biggr]^2 - 2 \biggl(
\frac{H}{\omega_{m}}\biggr)^2\Biggr\}.
\end{equation}
We are now ready to include the effect of the produced gravitational
waves in the Einstein equations.
The evolution of the Hubble parameter will be given (in the
conformally flat case)
by the following integro-differential equation:
\begin{eqnarray}
&&M^2_{P} H^2 - \rho_{s}(t) =  - \int_{t_{1}}^{t}
\biggl[\frac{a(t')}{a(t)}\biggr]^4
\biggl[ \frac{\partial
\overline{\rho}_{GW}(\omega_{m}(t'),t')}{\partial
\omega_{m}}\bigg]_{a} \frac{\partial \omega_{m}}{\partial t'} d
t',
\label{inte}\\
&&\rho_{s}(t) = H^2_1
M^2_{P}\biggl(\frac{a_1}{a}\biggr)^{3(\gamma+1)}
\nonumber\\
&&\biggl[\frac{\partial\overline{\rho}_{GW}}{\partial\omega_{m}}\biggr]_{a}
=\frac{\omega^3_{m}}{4~\pi^2}\Biggl\{ \biggl[\biggl(
\frac{a}{a_1}\biggr)-
\biggl(\frac{a_1}{a} \biggr)\biggr]^2 -
\frac{H^2}{\omega_m^2}\Biggr\}
\label{fin3}
\end{eqnarray}
In Eq. (\ref{fin3}) $\rho_{s}$
is just the energy density of the stiff background which decreases
faster that $a^{-4}$ for any $\gamma >1/3$. Notice that the right
hand
side of Eq. (\ref{inte}) is noting but the first of the FRW reported
in Eq. (\ref{FRW}). In the absence of graviton creation the left hand
side of Eq. (\ref{inte}) would be just zero, whereas in the presence
of graviton re-entering the horizon at any time after $t_{1}$ this
second term receives a  non vanishing contribution.

An useful way of re-writing the integral appearing in Eq.
(\ref{inte})
is by changing integration variable from $t'$ to $H'\equiv H(t')$,
and in this
way the total energy of produced gravitons becomes:
\begin{equation}
\rho_{{\rm tot}}(t) =
- \frac{1}{4 \pi^2}\int_{H_{1}}^{H}
\biggl[\frac{a(H')}{a(H)}\biggr]^4  \omega^3_{m}\Biggl\{ \biggl[
\frac{a}{a_1} - \frac{a_1}{a}\biggr]^2 - \biggl(
\frac{H'}{\omega_{m}}\biggr)^2 \Biggr\} dH'
\label{quasi}
\end{equation}
Notice that $\partial \omega_{m}/ \partial H \sim 1 $ since
$\omega_{m} \sim H$ when the given mode crosses the horizon.

It is clear that Eq. (\ref{fin3}) represents a complicated
integro-differential equation which cannot be solved exactly. One
possible way of dealing with this problem is to transform it into an
ordinary differential equation by solving the integral using the
scale factor of the stiff phase. In other words the integral at the
right hand side of Eq. (\ref{inte}) can be viewed as a perturbation
to
the solution of the FRW equations (\ref{FRW}) in the absence of
graviton creation.
Using this procedure we insert the stiff scale factor into
Eq. (\ref{fin3}) and we compute the effect of the gravitons
re-entering after $t_{1}$ on the background metric. Thus using
$[a(t')/a(t'_1)] \sim [t'/t_1]^{2/3(\gamma + 1)}$, $\omega_{m} \sim
1/t'$ and $H'=H(t')\sim 2/[3(\gamma +1) t']$ into Eq.  (\ref{quasi}),
our integro-differential equation becomes
\begin{eqnarray}
&&\biggl(\frac{dz}{dy}\biggr)^2 = e^{2y}\Biggl[
 \frac{4}{9(\gamma+1)^2} e^{-2 y} +\epsilon^2 e^{- 4x}
\Lambda(z)\Biggr]
\nonumber\\
&&\Lambda(z) =
\Biggl\{ f_{1}(\gamma)\biggl[e^{-2(1 + 3 \gamma)z} -1\biggr]
+f_{2}(\gamma)\biggl[1 -e^{-2(3 \gamma+2)z}\biggr]+
f_{3}(\gamma)\biggl[1 - e^{- 6\gamma z}\biggr]\Biggr\}
\end{eqnarray}
where
\begin{equation}
f_{1}(\gamma) = \frac{1}{24 \pi^2} \frac{ 2q +9 (1 + \gamma)}{(\gamma
+ 1) ( 3 \gamma + 1)^2},~~~f_{2}(\gamma) =
\frac{1}{16\pi^2} \frac{ 3 (\gamma + 1)}{2 + 3
\gamma},~~~f_{3}(\gamma) = \frac{1}{16\pi^2}
\biggl( \frac{\gamma + 1}{\gamma}\biggr),
\label{solve}
\end{equation}
(recall that $\epsilon = H_1/M_{P}$ \cite{staro} ).
 In Eq. (\ref{solve}) there is also the parameter
$q$ which needs to be explained. As we said before we regularized the
energy density using the adiabatic regularization scheme. Now, if we
set $q=1$ we automatically include in Eq. (\ref{solve}) the
subtractions coming from the adiabatic regularization. If we set $q
=0$ we practically use for the calculation of the energy density of
the gravitational waves the non-regularized energy density.
Thus`, if $q=1$,  $\overline{\rho}_{GW}$ is used in Eq. (\ref{fin3}).
If
$q=0$, Eq. (\ref{fin3}) is computed by using $\rho_{GW}$ which does
not
include the subtractions of the divergent terms. As we discussed
previously to set $q=1$ or $q=0$ does not change the numerical
solution which we are going to describe. It should be actually be
borne in mind that we have a physical cut-off provided by the class
of
models we are discussing and which is set by $k_1\sim \eta_{1}^{-1}$.
\begin{figure}
\begin{center}
\begin{tabular}{|c|c|}
      \hline
      \hbox{\epsfxsize = 7.5 cm  \epsffile{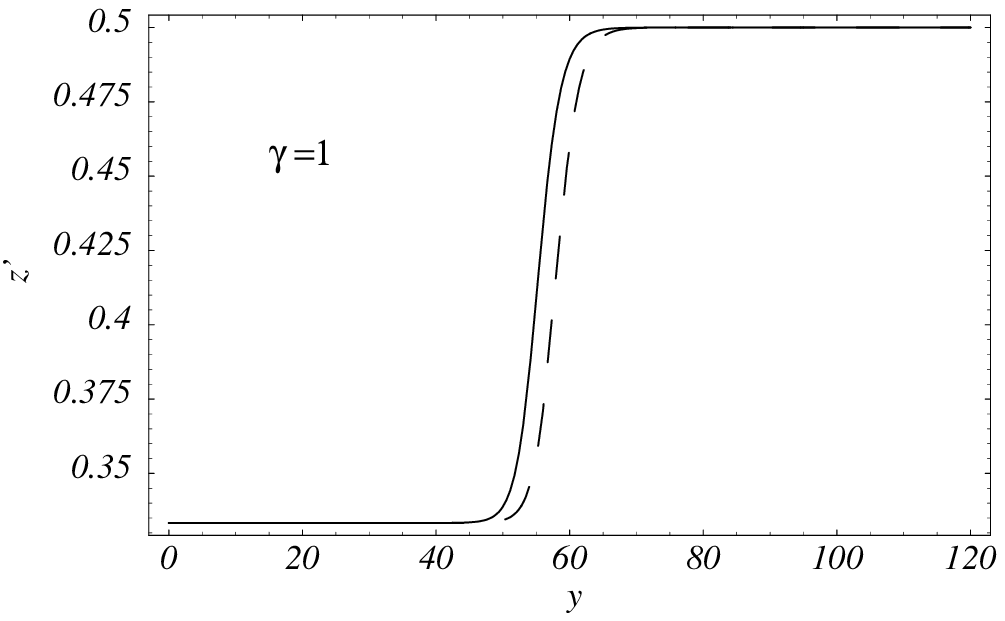}} &
      \hbox{\epsfxsize = 7.5 cm  \epsffile{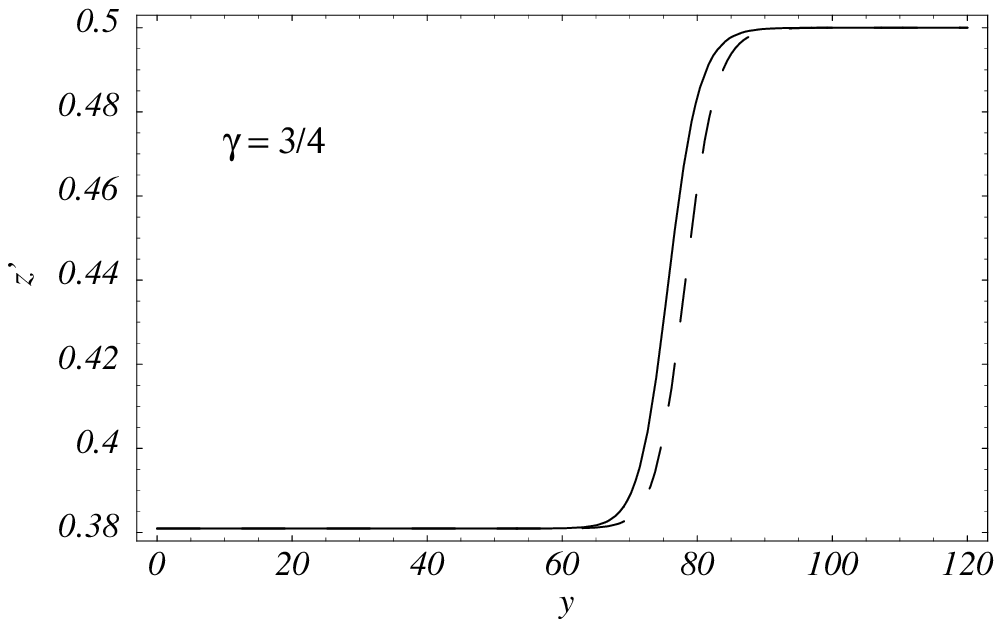}} \\
      \hline
      \hbox{\epsfxsize = 7.5 cm  \epsffile{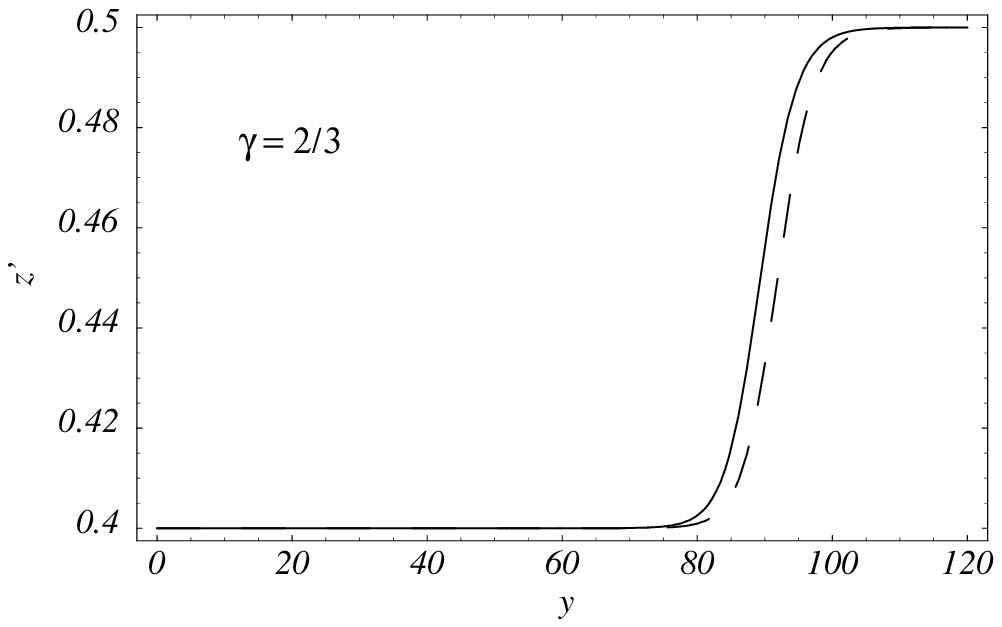}}  &
      \hbox{\epsfxsize = 7.5 cm  \epsffile{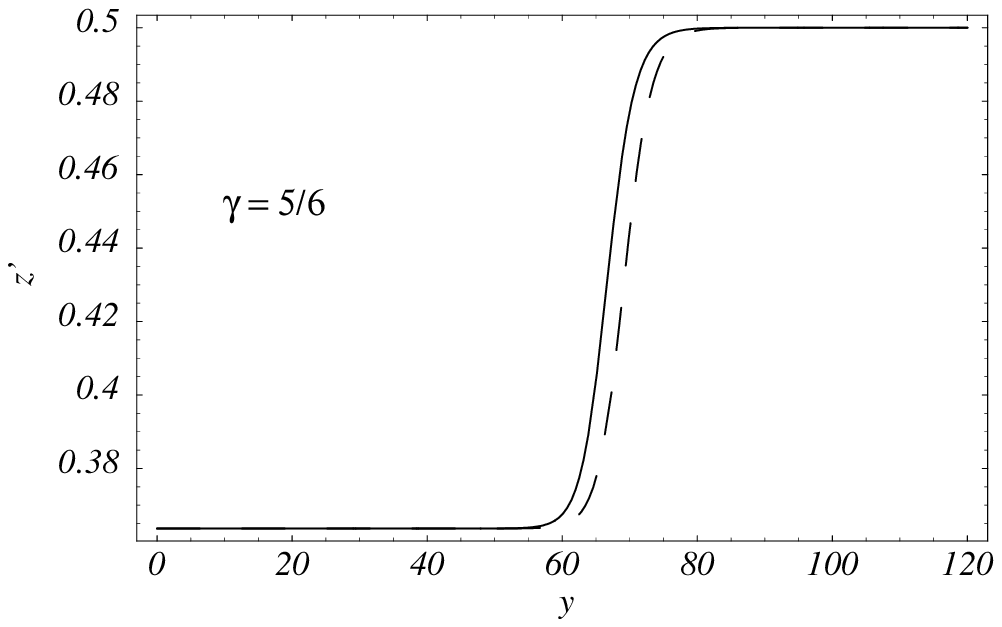}}\\
      \hline
\end{tabular}
\caption[a]{We plot $z'(y)$ in the case $\epsilon = 10^{-7}$. Recall
that
$z= \log{\frac{a}{a_1}}$ and $y= \log{\frac{t}{t_1}}$. Therefore, we
have that $\alpha(y)= z'(y)$. We plot the solutions for four
different
values of the initial $\gamma$ parameter. We integrated
Eq. (\ref{solve}) from $t=t_1$ (i.e.$y=0$) up to $y=120$.
Take, for instance, the case
$\gamma=1$ corresponding to $\alpha(t_1) = 1/3$. We see that, thanks
to the gravitational waves contribution the background undergoes a
phase transition towards a radiation dominated epoch with $z'(y) \sim
\alpha(y) \sim 1/2$. In the full lines is reported the case $q=0$
(no subtractions of the ultra-violet divergences) whereas in the full
lines is illustrated the case $q=1$ includeding the subtractions.}
\label{first}
\end{center}
\end{figure}
The effect of the subtractions (encoded in the choice of $q$) is
illustrated in Fig. \ref{first} where the solution of
Eq. (\ref{solve}) is reported as a function of $y=\log{t/t_1}$.
We can see that to include the subtractions affects the
transition regime (when the gravitational radiation starts to
dominate
the background) but not the asymptotic regime.
Notice that from the effective evolution of $\alpha(y)$ we can derive
also the effective evolution of $\gamma(y)$. In Fig. \ref{second1}
the
cosmic time evolution of $\gamma(y)$ is illustrated for the four
different cases of $\gamma(0)$ discussed in Fig. \ref{first}.
The back reaction which forces the
equation of state to pass from its original stiff value (for
$y\rightarrow 0$) to $1/3$ typical of a radiation dominated (perfect)
relativistic fluid.

\begin{figure}
\begin{center}
\begin{tabular}{|c|c|}
      \hline
      \hbox{\epsfxsize = 7.5 cm  \epsffile{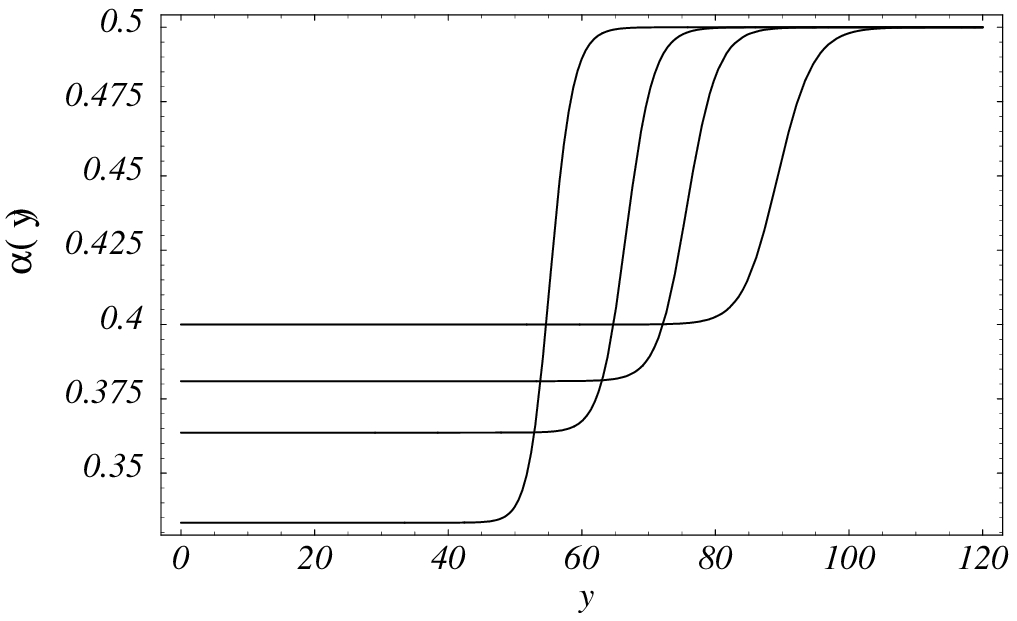}} &
      \hbox{\epsfxsize = 7.5 cm  \epsffile{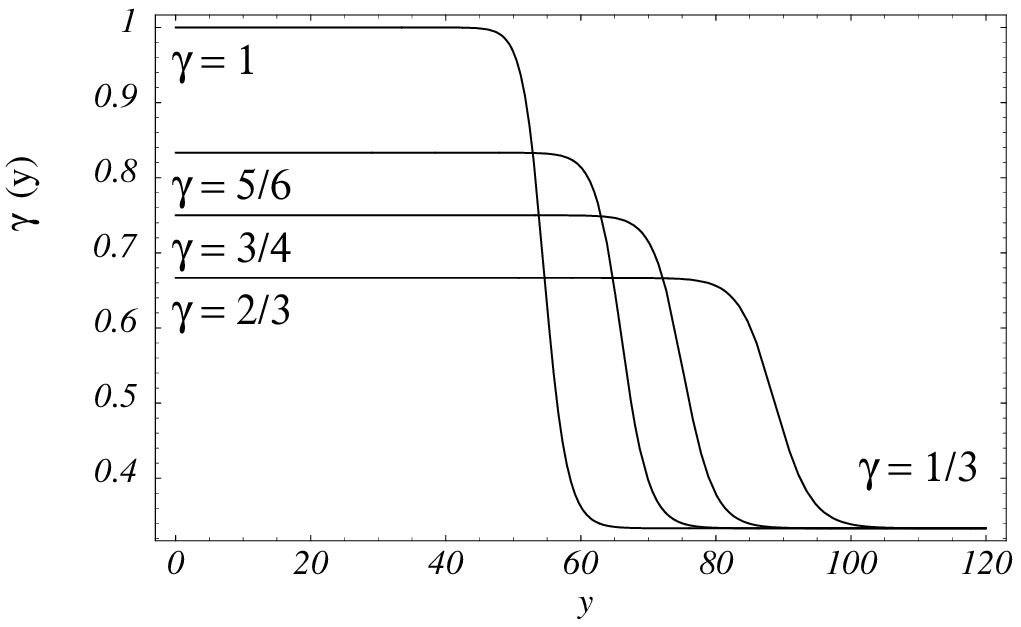}} \\
      \hline
\end{tabular}
\caption[a]{The time evolution of the equation of state is
reported. Recall, for comparison between the two pictures that
$\gamma(y) = 2/[3 z'(y)] -1$ and that $\alpha(y) = z'(y)$.
The four curves do correspond to the
four cases already discussed in Fig. \ref{first}. We notice that
starting with stiff equations of state at $t=t_1$ (i.e., from top to
bottom, $\gamma(0) =1,~5/6,~ 3/4,~2/3$) we are attracted towards
$\gamma
=1/3$ for large $y$.}
\label{second1}
\end{center}
\end{figure}
\begin{figure}
\begin{center}
\begin{tabular}{|c|c|}
      \hline
      \hbox{\epsfxsize = 7.5 cm  \epsffile{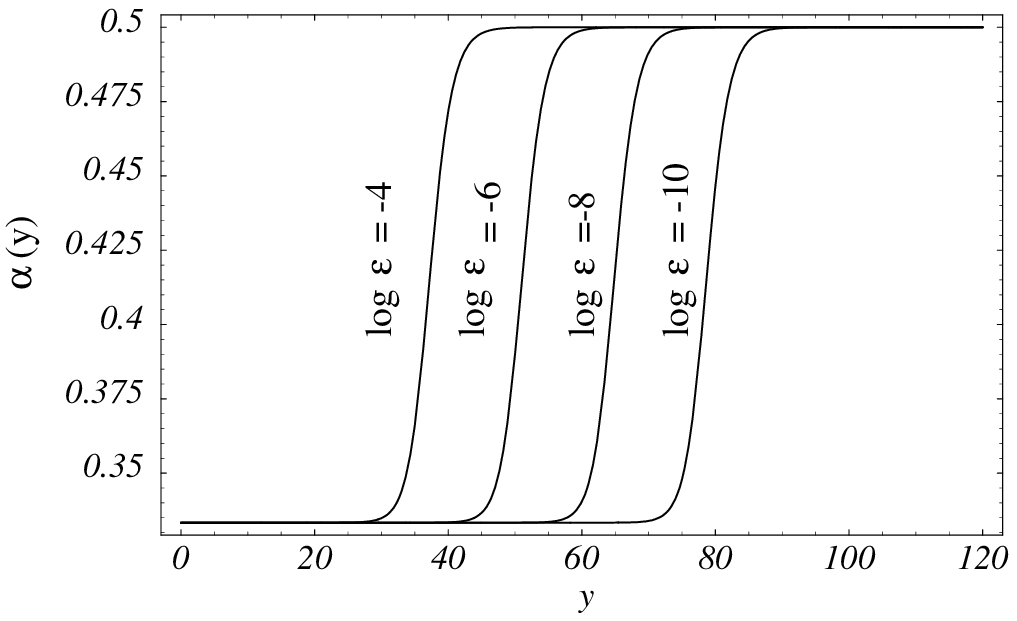}} &
      \hbox{\epsfxsize = 7.5 cm  \epsffile{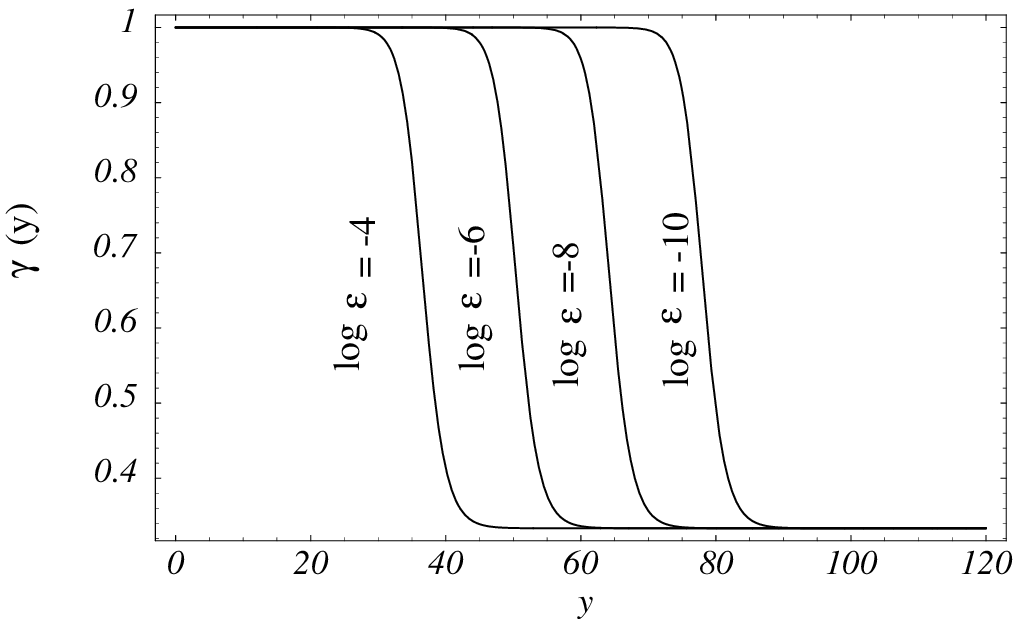}} \\
      \hline
\end{tabular}
\caption[a]{The solution of Eq. (\ref{solve}) is reported for the
case
$\gamma=1$. If $\epsilon= 10^{-4}$ (as required by BBN constraints)
we
have that the equation of state of the fluid driving the expansion
starts to deviate significantly from the stiff one already for
$y\sim 4~\times10^{9}$ corresponding to $H_{r}\sim 10^{-13} M_{P}$
(notice that always used neperian logarithms).}
\label{ep}
\end{center}
\end{figure}

It is also clear that, by lowering $\epsilon$ the transition to the
radiation dominated epoch might be delayed (paying, of course, the
price of a fine-tuning). In Fig \ref{ep} we report the solution of
Eq. (\ref{solve}) for different values of $\epsilon$ in the case
$\gamma=1$. As usual we can
either see the transition either in terms of $\gamma$ or in terms of
$\alpha$.

Looking at Fig. \ref{second1} and \ref{ep} we can also see that the
transition to the radiation dominated phase does not occur
instantaneously but it takes place in a finite amount of time which
turns out to be quite substantial (of the order of $20$ time
e-folding).

\renewcommand{\theequation}{5.\arabic{equation}}
\setcounter{equation}{0}
\section{Discussion and Conclusions}

In this paper we investigated the possibility of stiff epochs
occurring
immediately after an inflationary phase. We found that, if these
phases ever existed, they are constrained by the production of GW. We
computed the associated energy spectra of the produced gravitons and
we also discussed the associated back reaction effects.
For what concerns specifically the theoretical implications of GW
backgrounds of stiff origin, we can say that the possibility of
having blue spectra at high frequencies looks certainly promising. In
this class of models ``blue" spectra arise quite naturally. If the
maximal inflationary curvature scale is taken to be of the order of
$10^{-6} ~M_{P}$ we can also see from our results that the amplitude
of the GW (logarithmic) energy spectrum will be larger, at high
frequencies, than the inflationary prediction obtained in the absence
of stiff phases. The backreaction effect will make the Universe
dominated by radiation and the corresponding ``graviton reheating"
energy scale in the range  $1$--$10^{4}$ TeV.

The duration of the stiff phase can be long only if the maximal scale
where inflation occurs is fine-tuned to be much smaller than $10^{-6}
M_{P}$.
We also found that the transition regime (where the stiff equation of
state with $\gamma>1/3$) is replaced by $\gamma=1/3$ is quite long.
Our analysis has certainly different limitations. We mainly focused
our attention on the back reaction effects associated with high
frequency gravitons which behave effectively like radiation
\cite{grishchuk3,brandenberger,parker}.
The second limitation of our analysis is that we did not discuss the
amplification of the scalar fluctuations. We did not do this for the
simple reason that the scalar fluctuations are much more sensitive
to the particular dynamical model used in order to implement a stiff
phase.

A conservative conclusion of our investigation is that the the
duration of a stiff, post-inflationary phase is crucially determined
by the energy density of the inhomogeneity which were amplified
during
the inflationary phase and re-entered in the stiff phase.
In the framework of a particular model the back reaction effects of
tensor (and scalar) fluctuations should be analyzed.
On one hand the produced inhomogeneities could offer an original
mechanism for re-heating the Universe. On the other hand they can
forbid an arbitrary duration of the stiff phase.

\end{document}